\documentclass[12pt]{article}
\usepackage{amssymb,graphicx}

\begin{document}

\textheight 21.8cm

\newcommand{\sect}[1]{\setcounter{equation}{0}\section{#1}}
\renewcommand{\theequation}{\thesection.\arabic{equation}}
\newcommand{\be}{\begin{equation}}
\newcommand{\ee}{\end{equation}}
\newcommand{\bea}{\begin{eqnarray}}
\newcommand{\eea}{\end{eqnarray}}
\newcommand{\nonu}{\nonumber\\}
\newcommand{\beano}{\begin{eqnarray*}}
\newcommand{\eeano}{\end{eqnarray*}}
\newcommand{\eps}{\epsilon}
\newcommand{\om}{\omega}
\newcommand{\vph}{\varphi}
\newcommand{\sig}{\sigma}
\newcommand{\CC}{\mbox{${\mathbb C}$}}
\newcommand{\RR}{\mbox{${\mathbb R}$}}
\newcommand{\QQ}{\mbox{${\mathbb Q}$}}
\newcommand{\ZZ}{\mbox{${\mathbb Z}$}}
\newcommand{\NN}{\mbox{${\mathbb N}$}}
\newcommand{\1}{\mbox{\hspace{.0em}1\hspace{-.24em}I}}
\newcommand{\II}{\mbox{${\mathbb I}$}}
\newcommand{\prt}{\partial}
\newcommand{\und}[1]{\underline{#1}}
\newcommand{\wh}[1]{\widehat{#1}}
\newcommand{\wt}[1]{\widetilde{#1}}
\newcommand{\mb}[1]{\ \mbox{\ #1\ }\ }
\newcommand{\half}{\frac{1}{2}}
\newcommand{\noin}{\not\!\in}
\newcommand{\rhotimes}{\mbox{\raisebox{-1.2ex}{$\stackrel{\displaystyle\otimes}
{\mbox{\scriptsize{$\rho$}}}$}}}
\newcommand{\bin}[2]{{\left( {#1 \atop #2} \right)}}
\newcommand{\A}{{\cal A}}
\newcommand{\B}{{\cal B}}
\newcommand{\C}{{\cal C}}
\newcommand{\F}{{\cal F}}
\newcommand{\R}{{\cal R}}
\newcommand{\T}{{\cal T}}
\newcommand{\W}{{\cal W}}
\newcommand{\cS}{{\cal S}}
\newcommand{\bS}{{\bf S}}
\newcommand{\cL}{{\cal L}}
\newcommand{\hlp}{{\RR}_+}
\newcommand{\hlm}{{\RR}_-}
\newcommand{\Hil}{{\cal H}}
\newcommand{\D}{{\cal D}}
\newcommand{\bg}{{\bf g}}
\newcommand{\zf}{\A_{\cS}}
\newcommand{\alg}{\C_{\cS}}
\newcommand{\balg}{\B_{\cS}}
\newcommand{\malg}{\D_{S}}
\newcommand{\frep}{\F_{\R,\T}(\C_\cS)}
\newcommand{\frepb}{\F_{\R,\T}(\C_\B)}
\newcommand{\rep}{\F(\C_\cS)}
\newcommand{\ftrep}{{\cal G}(\C_\cS)}
\newcommand{\form}{\langle \, \cdot \, , \, \cdot \, \rangle }
\newcommand{\e}{{\rm e}}
\newcommand{\LL}{\mbox{${\mathbb L}$}}
\newcommand{\Rp}{{R^+_{\, \, \, \, }}}
\newcommand{\Rm}{{R^-_{\, \, \, \, }}}
\newcommand{\Rpm}{{R^\pm_{\, \, \, \, }}}
\newcommand{\Tp}{{T^+_{\, \, \, \, }}}
\newcommand{\Tm}{{T^-_{\, \, \, \, }}}
\newcommand{\Tpm}{{T^\pm_{\, \, \, \, }}}
\newcommand{\baral}{\bar{\alpha}}
\newcommand{\barbt}{\bar{\beta}}
\newcommand{\supp}{{\rm supp}\, }
\newcommand{\EE}{\mbox{${\mathbb E}$}}
\newcommand{\JJ}{\mbox{${\mathbb J}$}}
\newcommand{\MM}{\mbox{${\mathbb M}$}}
\newcommand{\ct}{{\cal T}}

\newtheorem{theo}{Theorem}[section]
\newtheorem{coro}[theo]{Corollary}
\newtheorem{prop}[theo]{Proposition}
\newtheorem{defi}[theo]{Definition}
\newtheorem{conj}[theo]{Conjecture}
\newtheorem{lem}[theo]{Lemma}
\newcommand{\prf}{\underline{\it Proof.}\ }
\newcommand{\finprf}{\null \hfill {\rule{5pt}{5pt}}\\[2.1ex]\indent}

\pagestyle{empty}
\rightline{December 2004}

\vfill

\begin{center}
{\Large\bf Reflection--Transmission Quantum Yang--Baxter Equations}
\\[2.1em]

{\large
V. Caudrelier$^{a}$\footnote{caudrelier@lapp.in2p3.fr}, 
M. Mintchev$^{b}$\footnote{mintchev@df.unipi.it},
E. Ragoucy$^{a}$\footnote{ragoucy@lapp.in2p3.fr}
and P. Sorba$^{a}$\footnote{sorba@lapp.in2p3.fr}}\\
\end{center}

\null

\noindent
{\it $^a$LAPTH, 9, Chemin de Bellevue, BP 110, F-74941 Annecy-le-Vieux
      cedex, France\\[2.1ex] 
$^b$INFN and Dipartimento di Fisica, Universit\'a di
      Pisa, Via Buonarroti 2, 56127 Pisa, Italy}
\vfill

\begin{abstract}

We explore the reflection--transmission quantum Yang--Baxter equations, arising 
in factorized scattering theory of integrable models with impurities. 
The physical origin of these equations is clarified and three general families of 
solutions are described in detail. Explicit representatives of each family are 
also displayed. These results allow to establish 
a direct relationship with the different previous works on the subject and make evident 
the advantages of the reflection--transmission algebra as an universal approach to 
integrable systems with impurities.

\end{abstract}

\vfill
\rightline{IFUP-TH 36/2004}
\rightline{LAPTH-1078/04}
\rightline{\tt hep-th/0412159}
\newpage
\pagestyle{plain}
\setcounter{page}{1}


\sect{Introduction} 

Quantum mechanics and quantum field theory with point-like impurities attract 
recently much attention in relation to the rapid progress of condensed matter 
physics with defects. Integrable systems with 
impurities represent in this context a relevant testing ground for the basic theoretical 
ideas and have also some direct physical applications. 

In the present Letter we are concerned with those universal features of integrable systems 
with impurities in 1+1 space-time dimensions, which are captured by 
the reflection-transmission quantum Yang-Baxter equations (QYBE's), following 
from factorized scattering theory. Our main goal below is to describe the 
origin and the present status of these equations. In the next section we sketch the derivation 
of the reflection--transmission QYBE's from first principles, namely physical unitarity 
and  the reflection (boundary) QYBE, familiar from the case of purely reflecting boundary. 
Afterwards we briefly describe the concept of reflection--transmission (RT) 
algebra, which is based on the reflection--transmission QYBE's. 
Sections 3 and 4 are devoted to some concrete examples of scattering data, which 
obey these equations and illustrate the general structure. 
In the last section we compare the existing two approaches to factorized 
scattering with impurities, displaying the advantages of the RT algebra framework. 
This section collects also our conclusions and indicates some further developments in the subject.   

\sect{Reflection-transmission QYBE's}

The method of factorized scattering \cite{Yang:1967bm} 
is a powerful tool for studying both the mathematical structure and the physical properties 
of integrable quantum systems in 1+1 dimensions. The main ingredient  
of this method is the two--body scattering matrix 
\be 
\{\cS_{\alpha_1 \alpha_2}^{\beta_1 \beta_2}(\chi_1,\chi_2) 
\, :\,  \alpha_1,...,\beta_2 = 1,...,N;\, \, \chi_1, \chi_2 \in \RR\} \, . 
\label{2body} 
\end{equation} 
Here $N$ is the number of internal 
degrees of freedom, whereas $\chi \in \RR$ paramet\-rizes the 
dispersion relation of the asymptotic particles. 
In order to construct from (\ref{2body}) a consistent 
total scattering operator, the  two-body matrix $\cS$ must satisfy  
\be
\cS_{12}(\chi_1,\chi_2) \cS_{13}(\chi_1,\chi_3) \cS_{23} (\chi_2,\chi_3)
= \cS_{23}(\chi_2,\chi_3) \cS_{13}(\chi_1,\chi_3) \cS_{12}(\chi_1,\chi_2)  \, ,   
\label{qyb}
\ee
which is the celebrated quantum Yang-Baxter equation (QYBE).  One can associate  
\cite{Sklyanin:ye}--\cite{Davies:gc} with $\cS$ an algebra $\zf$ with identity $\bf 1$, 
whose generators $\{a^{\ast \alpha}(\chi),\, a_\alpha (\chi)\}$  satisfy:   
\bea
a_{\alpha_1}(\chi_1) \, \, a_{\alpha_2 }(\chi_2) \, \; - \; \,
       \cS_{\alpha_2 \alpha_1 }^{\beta_2 \beta_1 }
       (\chi_2 , \chi_1)\,\, a_{\beta_2 }(\chi_2)\, a_{\beta_1 }(\chi_1) = 0
       \, , \qquad \quad \label{aa}\\
a^{\ast \alpha_1} (\chi_1)\, a^{\ast \alpha_2 } (\chi_2) -
       a^{\ast \beta_2 } (\chi_2)\, a^{\ast \beta_1 } (\chi_1)\,
       \cS_{\beta_2 \beta_1 }^{\alpha_2 \alpha_1 }(\chi_2 , \chi_1) = 0
       \, , \qquad \quad \label{a*a*} \\
a_{\alpha_1 }(\chi_1)a^{\ast \alpha_2 } (\chi_2)  - 
       a^{\ast \beta_2 }(\chi_2)\cS_{\alpha_1 \beta_2 }^{ \beta_1 \alpha_2}(\chi_1 , \chi_2) 
       a_{\beta_1 }(\chi_1) =   
       2\pi\delta (\chi_1 - \chi_2){\bf 1}  \, . 
\label{aa*}
\eea
The elements $\{a^{\ast \alpha}(\chi),\, a_\alpha (\chi)\}$ are interpreted as 
creators and annihilators of asymptotic particles. The QYBE (\ref{qyb}) ensures the 
associativity of $\zf$ and applying 
twice (\ref{aa}), one deduces the consistency relation 
\be
\cS_{12} (\chi_1,\chi_2)\, \cS_{21}(\chi_2,\chi_1) = \II\otimes \II \, ,
\label{unit}
\ee
known as {\it unitarity}. Moreover, requiring that the mapping 
\be
I\; :\; a^{\ast \alpha}(\chi) \mapsto a_\alpha (\chi)\, , \qquad 
I\; :\; a_\alpha(\chi) \mapsto a^{\ast \alpha} (\chi) \, , 
\label{involution}
\end{equation} 
generates an involution in $\zf$ (i.e. that $I$ extends as an antilinear anti\-homomorphism on $\zf$), 
one gets the so called {\it Hermitian analyticity} condition
\be
\cS_{12}^\dagger (\chi_1,\chi_2) = \cS_{21}(\chi_2,\chi_1)\, ,
\label{ha}
\ee
where the dagger stands for the Hermitian conjugation.\footnote {More general involutions in $\zf$ and the 
relative Hermitian analyticity conditions have been studied in \cite{Liguori:de}.} We stress that 
combining (\ref{unit}) and (\ref{ha}), which are assumed throughout the paper, 
one deduces the {\it physical} unitarity 
\be
\cS_{12} (\chi_1,\chi_2) \cS_{12}^\dagger (\chi_1,\chi_2)= \II\otimes \II   
\label{physunit}
\ee
of the two-body scattering matrix. Following the already standard terminology, 
in what follows we refer to $\zf$ as Zamolodchikov-Faddeev (ZF) algebra. 

The above framework has been successfully generalized \cite{Cherednik:vs}-\cite{Fring:1993mp}
to the case when a purely reflecting boundary is present in the space. 
Describing the process of particle reflection from the boundary 
by a reflection matrix $\R_\alpha^\beta (\chi)$, Cherednik \cite{Cherednik:vs} discovered 
in the early eighties that $\R$ must satisfy the following {\it reflection} QYBE 
\bea
\cS_{12}(\chi_1 , \chi_2)\, \R_1(\chi_1)\,
\cS_{21}(\chi_2 , -\chi_1)\, \R_2(\chi_2) = \nonumber\\
\R_2(\chi_2)\, \cS_{12}(\chi_1 , -\chi_2)\, \R_1(\chi_1)\,
\cS_{21}(-\chi_2 , -\chi_1)\, 
\label{rqyb}
\eea 
in order to have consistent factorized scattering. 
In analogy with (\ref{unit},\ref{ha}) one requires also unitarity 
\be 
\R(\chi) \R(-\chi) = \II 
\label{runit}
\end{equation} 
and Hermitian analyticity 
\be 
\R^\dagger(\chi)= \R(-\chi) \, , 
\label{rha}
\end{equation} 
which imply the physical unitarity 
\be 
\R(\chi) \R^\dagger(\chi) = \II \, . 
\label{rpunit}
\end{equation} 
Let us mention in passing that the ZF algebra $\zf$ has a counterpart $\balg$ 
in the boundary case \cite{Liguori:1998xr}. Instead of describing $\balg$ now, we will obtain it later on 
as a special case of the more general structure, discussed below. 

It is quite natural at this stage to consider instead of the purely reflecting boundary 
an impurity (defect), which both reflects and transmits.  In addition to 
$\R_\alpha^\beta (\chi)$, one will have in this case also a transmission matrix 
$\T_\alpha^\beta(\chi)$. Quantum mechanical potential scattering 
theory (see e.g. \cite{A}) suggests to substitute (\ref{runit}) by 
\bea
\T(\chi) \T(\chi) + \R(\chi) \R(-\chi) = \II\, , 
\label{rtunit1} \\ 
\T(\chi) \R(\chi) + \R(\chi) \T(-\chi) = 0 \, ,
\label{rtunit2}
\eea
where $\T$ satisfies the Hermitian analyticity condition 
\be 
\T(\chi) = \T^\dagger (\chi ) \, . 
\label{tha}
\end{equation}
Due to (\ref{rha}), (\ref{tha}), $\T(\chi) \T(\chi)$ and $\R (\chi) 
\R (-\chi)$ are non--negative Hermitian matrices which are
simultaneously diagonalizable because of (\ref{rtunit1}). 
The corresponding eigenvalues $\lambda_i(\chi)$ and $\mu_i(\chi)$ satisfy
\be
0\leq \lambda_i(\chi) \leq 1\, ,
\quad 0\leq \mu_i(\chi) \leq 1\, , \quad 
\lambda_i(\chi) + \mu_i(\chi) = 1\, ,
\qquad i=1,...,N.
\label{eig}
\end{equation}
Obviously, non-trivial transmission occurs if and only if 
$\mu_i(\chi) <1$ (or equivalently $\lambda_i(\chi) >0$) for some $i=1,...,N$. 

{}Following \cite{Mintchev:2002zd, Mintchev:2003ue}, our goal now 
is to uncover the algebraic structure which models  
the mechanism of reflection and transmission in integrable systems with 
impurities. It is natural to expect that in addition to Cherednik's 
reflection QYBE (\ref{rqyb}), $\R$ and $\T$ satisfy some 
{\it transmission} and {\it reflection-transmission} QYBE's as well. 
The most direct way to derive these equations is to solve the unitarity constraints 
(\ref{rtunit1}, \ref{rtunit2}), expressing $\T$ in terms of 
$\R$, and use afterwards (\ref{rqyb}). Solving (\ref{rtunit1}), we consider below 
the positive square root $\sqrt{\II - \R(\chi) \R(-\chi)}$ of the non-negative Hermitian matrix 
$\II - \R(\chi) \R(-\chi)$. More precisely, in the basis in which $\II - \R(\chi) \R(-\chi)$ is 
diagonal we take diagonal $\sqrt{\II - \R(\chi) \R(-\chi)}$, whose elements are 
the positive square roots $\sqrt{1-\lambda_i}$. The latter are well-defined in view of 
(\ref{eig}) and we set 
\be
\T(\chi) = \tau(\chi) \sqrt{\II - \R(\chi) \R(-\chi)} \, ,
\label{expT}
\end{equation}
where the $\tau$--function obeys 
\be
{\overline \tau}(\chi) = \tau (\chi)\, , \qquad \tau(\chi)^2 = 1\, ,
\qquad \tau(-\chi) = - \tau(\chi) \, , 
\label{tau}
\end{equation}
following from (\ref{rtunit1}--\ref{tha}). If one assumes in addition that $\tau$ is 
continuous for $\chi\not= 0$, one finds 
\be 
\tau (\chi) = \pm \varepsilon (\chi ) \, , 
\label{H}
\end{equation} 
$\varepsilon$ being the sign function. Adopting the representation (\ref{expT}) and the 
reflection QYBE (\ref{rqyb}), one can prove \cite{Mintchev:2003ue} the following statement. 

\bigskip 

\noindent{\bf Proposition:} {\it Let $\R$ and $\T$ satisfy hermitian analyticity} (\ref{rha},\ref{tha}), 
{\it unitarity} (\ref{rtunit1}, \ref{rtunit2}) {\it and the reflection QYBE} (\ref{rqyb}). {\it Then $\T$ 
defined by} (\ref{expT}) {\it and $\R$ obey the transmission QYBE 
\bea
\cS_{12}(\chi_1 , \chi_2)\, \T_1(\chi_1)\,
\cS_{21}(\chi_2 , \chi_1)\, \T_2(\chi_2) = \nonumber\\
\T_2(\chi_2)\, \cS_{12}(\chi_1 , \chi_2)\, \T_1(\chi_1)\,
\cS_{21}(\chi_2 , \chi_1)\, 
\label{tqyb}
\eea
and the reflection-transmission QYBE
\bea
\cS_{12}(\chi_1 , \chi_2)\, \T_1(\chi_1)\,
\cS_{21}(\chi_2 , \chi_1)\, \R_2(\chi_2) = \nonumber\\
\R_2(\chi_2)\, \cS_{12}(\chi_1 , -\chi_2)\,
\T_1(\chi_1)\, \cS_{21} (-\chi_2 , \chi_1)\,  
\label{rtqyb}
\eea
as well}. 
\bigskip 

\noindent {\bf Remark}: The precise determination of the square root we are taking in (\ref{expT})  
is essential. The point is that $\II - \R(\chi) \R(-\chi)$ has in general infinitely 
many square roots - a standard phenomenon \cite {G, H} in matrix theory. 
By investigating some concrete examples we have seen that instead of 
eqs. (\ref{tqyb}, \ref{rtqyb}), some of the roots obey more complicated ``twisted" versions 
of these equations in which $\T$ is substituted by $A\T A^{-1}$ with  
an invertible matrix $A$. Postponing the study of the latter case to the future, 
we focus in this paper on the subset of roots  satisfying (\ref{tqyb}, \ref{rtqyb}). 
The above proposition simply states that $\T$, corresponding to the positive 
square root of $\II - \R(\chi) \R(-\chi)$, belongs to this subset.  

Equations (\ref{tqyb}, \ref{rtqyb}) have the same structure as (\ref{rqyb}), but for 
some sign-changes in the arguments of $\cS$ with obvious 
kinematical interpretation. They are essential for the reconstruction \cite{Mintchev:2003ue} of the 
reflection-transmission (RT) algebra $\alg$ - the analog of $\zf$ when impurities are present. 
It turns out that in addition to $\{a^{\ast \alpha}(\chi),\, a_\alpha (\chi)\}$, 
$\alg$ involves $2N^2$ defect generators 
$\{t_\alpha^\beta (\chi),\, r_\alpha^\beta (\chi)\}$, which describe the particle 
interaction with the impurity and modify the right hand side of the exchange relation (\ref{aa*}) as follows 
\bea
a_{\alpha_1 }(\chi_1)\, a^{\ast \alpha_2 } (\chi_2) \; - \;
       a^{\ast \beta_2 }(\chi_2)\,
       \cS_{\alpha_1 \beta_2 }^{\beta_1 \alpha_2}(\chi_1 , \chi_2)\,
       a_{\beta_1 }(\chi_1) =  \nonumber\\
       2\pi\, \delta (\chi_1 - \chi_2)\,
       \left [\delta_{\alpha_1 }^{\alpha_2 }\, {\bf 1} +
t_{\alpha_1}^{\beta_2}(\chi_1)\right] +
       2\pi\, \delta (\chi_1 + \chi_2)\,  r_{\alpha_1 }^{\alpha_2 }(\chi_1)
       \, . 
\label{rtaa*}
\eea 
The relations (\ref{aa}, \ref{a*a*}) remain invariant. Finally, for defining $\alg$ one 
must add the exchange relations among $\{t_\alpha^\beta (\chi),\, r_\alpha^\beta (\chi)\}$ 
themselves and with\break $\{a^{\ast \alpha}(\chi),\, a_\alpha (\chi)\}$. Since these relations will be 
of no use in this paper, we omit them, referring for the explicit form of the 
complete set of constraints imposed on the generators of $\alg$ to \cite{Mintchev:2003ue}. 
The boundary algebra $\balg$, mentioned above, is obtained from $\alg$ by setting 
$t_{\alpha}^{\beta}(\chi) = 0$. The interplay between the three algebras $\zf$, $\balg$ and $\alg$ 
has been investigated in \cite{Mintchev:2003kh, Ragoucy:2001cf}. 

$\alg$ is an infinite algebra and from its formal definition
it is not obvious at all that it has an operator realization. For this reason 
the Fock representation $\rep$ has been constructed in \cite{Mintchev:2003ue} explicitly in terms 
of (generally unbounded) operators acting on a suitable dense domain of a Hilbert space.  
In the construction of $\rep$ one needs an involution in $\alg$, which is obtained by extending 
(\ref{involution}) to the reflection and transmission generators according to   
\be
I \, : \,  r_\alpha^\beta  (\chi ) \mapsto r_\beta^\alpha (-\chi )\,
, \qquad \quad \, \, \,
I \, : \,  t_\alpha^\beta  (\chi ) \mapsto t_\beta^\alpha (\chi )\, . 
\label{inv2}
\ee 
Hermitian analyticity (\ref{rha}, \ref{tha}) ensures the consistency of this extension. 
We would like to recall also that $\{t_\alpha^\beta (\chi),\, r_\alpha^\beta (\chi)\}$ condense 
in the vacuum state $\Omega \in \rep$, i.e.   
\be 
\langle t_\alpha^\beta (\chi) \rangle_{{}_\Omega} = \T_\alpha^\beta (\chi)\, , \qquad 
\langle r_\alpha^\beta (\chi) \rangle_{{}_\Omega} = \R_\alpha^\beta (\chi)\, . 
\label{condensates}
\end{equation} 
This direct relationship between reflection--transmission generators and amplitudes is 
crucial for the physical interpretation.  

The Fock representation $\rep$ is useful in several respects. From 
the scattering data $\{\cS,\, \R,\, \T\}$ one can reconstruct \cite{Mintchev:2003ue} in $\rep$ the 
total scattering operator, which is unitary as expected. Remarkably enough, 
$\rep$ applies also in the construction of off--shell interacting quantum fields. 
An instructive example in this respect is the non--linear Schr\"odinger (NLS) model 
(the non--relativistic $\varphi^4$--theory) with 
a point--like impurity in 1+1 dimensions. This system is investigated in \cite{Caudrelier:2004gd, Caudrelier:2004xy}, 
where the exact off--shell operator solution is constructed in terms of the representation 
$\rep$ of an appropriate RT algebra $\alg$. Recently, this solution has been generalized \cite{CR} 
to the case of the $GL(N)$--invariant NLS model, where inequivalent Fock representations 
$\rep$ implement a mechanism of spontaneous symmetry breaking. 

Let us mention finally that $\alg$ admits also finite temperature representations. 
In contrast to the Fock vacuum, the cyclic state of such representations 
is not annihilated by $a_\alpha (\chi)$ and models a thermal bath, keeping the 
system in equilibrium at fixed (inverse) temperature $\beta$. 
A finite temperature representation of $\alg$ is introduced in \cite{Mintchev:2004jy}, 
where also some applications to the statistical mechanics of systems with impurities 
are discussed. 

Summarizing, the RT algebra admissible scattering data $\{\cS,\, \R,\, \T\}$ must satisfy 
the QYBE's (\ref{qyb}, \ref{rqyb}, \ref{tqyb}, \ref{rtqyb}), 
unitarity (\ref{unit}, \ref{rtunit1}, \ref{rtunit2}) and 
hermitian analyticity (\ref{ha}, \ref{rha}, \ref{tha}). The relevant issue at this point is 
the classification of admissible triplets. This is a hard theoretical problem, 
which has not been yet solved even in the case  of the QYBE (\ref{qyb}) alone. 
Fortunately however, some classes of admissible scattering data $\{\cS,\, \R,\, \T\}$ are 
known. They are the subject of our discussion in the next sections. 

\bigskip 

\sect{RT algebra admissible data $\{\cS,\, \R,\, \T\}$} 

Let us fix first of all the physical setting.  We will study a system with a single impurity in 
$\RR$. Since the impurity divides $\RR$ in two disconnected parts $\RR_\pm$, 
we take $N=2n$ and split the index $\alpha$ as follows $\alpha = (\xi, i)$. Here  
$\xi = \pm$ indicates the half--line 
where the asymptotic particle is created or annihilated and $i=1,...,n$ labels the 
``isotopic" type. We emphasize that the results of the present paper are valid  for any 
dispersion relation 
\be 
E=E(\chi)\, , \qquad p = p(\chi) \, , 
\label{dr}
\end{equation} 
between the particle energy $E$ and momentum $p$. In particular, it is instructive to keep in mind
the conventional relativistic 
\be
E(\chi) = m \cosh (\chi)\, , \qquad p(\chi) = m \sinh (\chi)\, ,
\label{rel}
\ee
and non-relativistic
\be
E(\chi) = \frac{m\chi^2}{2} + U\, , \qquad p(\chi) = m\chi \, ,
\label{nrel}
\ee
relations, $m$ being the particle mass and $U$ some constant.  We observe that Lorentz and Galilean 
transformations of the vector $(E,p)$ are implemented in both (\ref{rel}) and (\ref{nrel})
by translations $\chi \mapsto \chi + \alpha$. Therefore the scattering matrix $\cS$ is 
Lorentz (Galilean) invariant provided that it depends on $\chi_1$ and $\chi_2$ only through 
the difference $\chi_1-\chi_2$.  

We focus in this section on block--diagonal $\cS$-matrices 
\begin{eqnarray}
    \cS(\chi_1,\chi_2) = \left(\begin{array}{cccc}
    S^{++}(\chi_1,\chi_2) & 0 & 0 & 0\\
    0 & S^{+-}(\chi_1,\chi_2) & 0 & 0\\
    0 & 0 & S^{-+}(\chi_1,\chi_2) & 0\\
    0 & 0 & 0 & S^{--}(\chi_1,\chi_2) 
    \end{array}\right)\, ,\nonumber \\
{}\label{smat}
\end{eqnarray}
each block being a $n^2\times n^2$ matrix in the isotopic space. The QYBE for $\cS$ generates 
six equations for its blocks.  One has 
\be
S^{++}_{\, 12}(\chi_1,\chi_2)\, S^{++}_{\, 13}(\chi_1,\chi_3)\, S^{++}_{\, 23} (\chi_2,\chi_3) = 
S^{++}_{\, 23} (\chi_2,\chi_3)\, S^{++}_{\, 13}(\chi_1,\chi_3)\, S^{++}_{\, 12}(\chi_1,\chi_2)  \, ,  
\label{qyb1}
\end{equation}
\be
S^{++}_{\, 12}(\chi_1,\chi_2)\, S^{+-}_{\, 13}(\chi_1,\chi_3)\, S^{+-}_{\, 23} (\chi_2,\chi_3) = 
S^{+-}_{\, 23} (\chi_2,\chi_3)\, S^{+-}_{\, 13}(\chi_1,\chi_3)\, S^{++}_{\, 12}(\chi_1,\chi_2)  \, ,  
\label{qyb2}
\end{equation}
\be
S^{+-}_{\, 12}(\chi_1,\chi_2)\, S^{++}_{\, 13}(\chi_1,\chi_3)\, S^{-+}_{\, 23} (\chi_2,\chi_3) = 
S^{-+}_{\, 23} (\chi_2,\chi_3)\, S^{++}_{\, 13}(\chi_1,\chi_3)\, S^{+-}_{\, 12}(\chi_1,\chi_2)  \, , 
\label{qyb3}
\end{equation}
\be
S^{+-}_{\, 12}(\chi_1,\chi_2)\, S^{+-}_{\, 13}(\chi_1,\chi_3)\, S^{--}_{\, 23} (\chi_2,\chi_3) = 
S^{--}_{\, 23} (\chi_2,\chi_3)\, S^{+-}_{\, 13}(\chi_1,\chi_3)\, S^{+-}_{\, 12}(\chi_1,\chi_2)  \, , 
\label{qyb4}
\end{equation}
the remaining four equations following from (\ref{qyb1}-\ref{qyb4}) with the exchange\break 
$+ \leftrightarrow -$. 
We will call the latter the mirror counterparts of (\ref{qyb1}-\ref{qyb4}). 
All of these equations have the structure of a QYBE. We stress however that apart from 
(\ref{qyb1}) and its mirror counterpart, the other four equations involve different matrices in 
the isotopic space and are therefore not genuine QYBE's. Similar equations appear in 
the context of the quartic algebras introduced by Freidel and Maillet \cite{Freidel:1991jx}. 

Taking $\R$ and $\T$ of the form 
\begin{eqnarray}
    \R(\chi) = \left(\begin{array}{cccc}
    R^+(\chi ) & 0\\
    0 & R^-(\chi) 
    \end{array}\right)\,  , \qquad  
\T(\chi) = \left(\begin{array}{cccc}
    0 & T^+(\chi )\\
    T^-(\chi ) & 0 
    \end{array}\right)\,  , 
\label{rtmat}
\end{eqnarray}
Hermitian analyticity (\ref{rha}, \ref{tha}) and unitarity (\ref{rtunit1}, \ref{rtunit2}) imply
\be 
\left [R^\pm\right ]^\dagger (\chi )= R^\pm (-\chi )\, , \qquad 
\left [T^+\right ]^\dagger (\chi ) = T^- (\chi )\, , 
\label{ha3}
\end{equation}
and 
\bea
T^\pm(\chi) T^\mp (\chi) + R^\pm (\chi) R^\pm (-\chi) = \II\, , 
\label{rtunit3} \\ 
T^\pm(\chi) R^\mp(\chi) + R^\pm(\chi) T^\pm(-\chi) = 0 \, ,
\label{rtunit4}
\eea
respectively. Moreover, from the QYBE's  (\ref{rqyb}, \ref{tqyb}, \ref{rtqyb}) 
one infers  
\bea
S^{++}_{\, 12}(\chi_1,\chi_2)\, R^+_1(\chi_1)\, S^{++}_{\, 21}(\chi_2,-\chi_1)\, R^+_2(\chi_2) = 
\nonumber \\
R^+_2(\chi_2)\, S^{++}_{\, 12}(\chi_1,-\chi_2)\, R^+_1(\chi_1)\, S^{++}_{\, 21}(-\chi_2,-\chi_1) 
\, , 
\label{rqyb1}
\eea 
\bea
S^{+-}_{\, 12}(\chi_1,\chi_2)\, R^+_1(\chi_1)\, S^{-+}_{\, 21}(\chi_2,-\chi_1)\, R^-_2(\chi_2) = 
\nonumber \\
R^-_2(\chi_2)\, S^{+-}_{\, 12}(\chi_1,-\chi_2)\, R^+_1(\chi_1)\, S^{-+}_{\, 21}(-\chi_2,-\chi_1) \, , 
\label{rqyb2}
\eea 
\bea
S^{++}_{\, 12}(\chi_1,\chi_2)\, T^+_1(\chi_1)\, S^{+-}_{\, 21}(\chi_2,\chi_1)\, T^+_2(\chi_2) = 
\nonumber \\
T^+_2(\chi_2)\, S^{+-}_{\, 12}(\chi_1,\chi_2)\, T^+_1(\chi_1)\, S^{--}_{\, 21}(\chi_2,\chi_1) 
\, , 
\label{tqyb1}
\eea 
\bea
S^{+-}_{\, 12}(\chi_1,\chi_2)\, T^+_1(\chi_1)\, S^{--}_{\,  21}(\chi_2,\chi_1)\, T^-_2(\chi_2) = 
\nonumber \\
T^-_2(\chi_2)\, S^{++}_{\,  12}(\chi_1,\chi_2)\, T^+_1(\chi_1)\, S^{+-}_{\,  21}(\chi_2,\chi_1) \, ,  
\label{tqyb2}
\eea 
\bea
S^{++}_{\,  12}(\chi_1,\chi_2)\, T^+_1(\chi_1)\, S^{+-}_{\,  21}(\chi_2,\chi_1)\, R^+_2(\chi_2) = 
\nonumber \\
R^+_2(\chi_2)\, S^{++}_{\,  12}(\chi_1,-\chi_2)\, T^+_1(\chi_1)\, S^{+-}_{\,  21}(-\chi_2,\chi_1) 
\, , 
\label{rtqyb1}
\eea 
\bea
S^{+-}_{\,  12}(\chi_1,\chi_2)\, T^+_1(\chi_1)\, S^{--}_{\,  21}(\chi_2,\chi_1)\, R^-_2(\chi_2) = 
\nonumber \\
R^-_2(\chi_2)\, S^{+-}_{\,  12}(\chi_1,-\chi_2)\, T^+_1(\chi_1)\, S^{--}_{\,  21}(-\chi_2,\chi_1)  
\label{rtqyb2}
\eea 
and their mirror analogs obtained by $+ \leftrightarrow -$ in (\ref{rqyb1}--\ref{rtqyb2}). 
Some solutions of the above equations are described in the next section. 

In order to keep the discussion as simple as possible, 
we concentrated above on triplets $\{\cS,\, \R,\, \T\}$ 
of the form (\ref{smat}, \ref{rtmat}), but more general admissible scattering 
data can be analyzed along the same lines.

\sect{Some families of scattering data}

Different classes of solutions of eqs. (\ref{qyb1}-\ref{qyb4}) determine different 
families of admissible scattering data. 
We concentrate on three of them, which proved to be
fundamental in the understanding of the physical content of RT
algebras. Indeed, the type I family answers the long-standing
question of the relationship between RT algebras and the approach 
to impurities in integrable systems followed in 
\cite{Cherednik:1990ju}-\cite{Delfino:1994nr}. 
It shows that the RT algebra framework is more general and
reproduces the  content of \cite{Cherednik:1990ju}-\cite{Delfino:1994nr} 
as a very special case. This is at the heart of the discussion of
the next section. Our concern with the type II family is to
exhibit a case where one can implement Lorentz invariance in the 
bulk scattering matrix, keeping non-trivial transmission. 
After presenting the general family,
we show an explicit example where this is realized. Finally, the
type III family is an important class of data since it involves 
the first example of interacting, exactly solvable and 
integrable quantum field model with impurity \cite{Caudrelier:2004gd}-\cite{CR}.

We turn now to the detailed describtion of the families I-III. 

\bigskip 
(i) {\bf Type I} scattering data are determined by 
\bea 
S^{+-}(\chi_1,\chi_2) = S^{-+}(\chi_1,\chi_2) = \II \, ,  \qquad \qquad 
\nonumber \\
S^{++}(\chi_1,\chi_2) =  S(\chi_1, \chi_2) \,  ,\qquad 
S^{--}(\chi_1,\chi_2) = {\widetilde S}(\chi_1, \chi_2)\, , 
\label{IS}
\end{eqnarray}
where $S$ and ${\widetilde S}$ are in general two different solutions of the QYBE. 
Bulk scattering matrices of the type (\ref{IS}) appear in \cite{Bajnok:2004jd}, 
where the so called ``folding trick" for describing reflecting and 
transmitting impurities is attempted. It is instructive to see what are the implications of 
eqs. (\ref{rqyb1}--\ref{rtqyb2}) and their mirror counterparts on $\R$ and $\T$. One finds 
the following reflection 
\bea
S_{12}(\chi_1,\chi_2)\, R^+_1(\chi_1)\, S_{21}(\chi_2,-\chi_1)\, R^+_2(\chi_2) = 
\nonumber \\
R^+_2(\chi_2)\, S_{12}(\chi_1,-\chi_2)\, R^+_1(\chi_1)\, S_{21}(-\chi_2,-\chi_1) \, , 
\label{rqyb11}
\eea 
\bea
{\widetilde S}_{12}(\chi_1,\chi_2)\, R^-_1(\chi_1)\, {\widetilde S}_{21}(\chi_2,-\chi_1)\, R^-_2(\chi_2) = 
\nonumber \\
R^-_2(\chi_2)\, {\widetilde S}_{12}(\chi_1,-\chi_2)\, R^-_1(\chi_1)\, {\widetilde S}_{21}(-\chi_2,-\chi_1) \, , 
\label{rqyb22}
\eea
transmission 
\be
T^+_1(\chi_1)\, {\widetilde S}_{21}(\chi_2,\chi_1)\, T^-_2(\chi_2) = 
\nonumber \\
T^-_2(\chi_2)\, S_{12}(\chi_1,\chi_2)\, T^+_1(\chi_1) \, ,  
\label{tqyb11}
\end{equation} 
\be
S_{12}(\chi_1,\chi_2)\, T^+_1(\chi_1)\, T^+_2(\chi_2) = 
T^+_2(\chi_2)\, T^+_1(\chi_1)\, {\widetilde S}_{21}(\chi_2,\chi_1) \, , 
\label{tqyb22}
\end{equation} 
\be
{\widetilde S}_{12}(\chi_1,\chi_2)\, T^-_1(\chi_1)\, T^-_2(\chi_2) = 
T^-_2(\chi_2)\, T^-_1(\chi_1)\, S_{21}(\chi_2,\chi_1) \, , 
\label{tqyb33}
\end{equation} 
and reflection--transmission equations
\be
S_{12}(\chi_1,\chi_2)\, T^+_1(\chi_1)\, R^+_2(\chi_2) = 
\nonumber \\
R^+_2(\chi_2)\, S_{12}(\chi_1,-\chi_2)\, T^+_1(\chi_1)\, , 
\label{rtqyb11}
\end{equation}
\be
T^+_1(\chi_1)\, {\widetilde S}_{\,  21}(\chi_2,\chi_1)\, R^-_2(\chi_2) = 
\nonumber \\
R^-_2(\chi_2)\, T^+_1(\chi_1)\, {\widetilde S}_{21}(-\chi_2,\chi_1)\, ,  
\label{rtqyb22}
\end{equation}
\be
{\widetilde S}_{\,  12}(\chi_1,\chi_2)\, T^-_1(\chi_1)\, R^-_2(\chi_2) = 
\nonumber \\
R^-_2(\chi_2)\, {\widetilde S}_{12}(\chi_1,-\chi_2)\, T^-_1(\chi_1) \, , 
\label{rtqyb33}
\end{equation}
\be
T^-_1(\chi_1)\, S_{21}(\chi_2,\chi_1)\, R^+_2(\chi_2) = 
\nonumber \\
R^+_2(\chi_2)\, T^-_1(\chi_1)\, S_{21}(-\chi_2,\chi_1)\, .  
\label{rtqyb44}
\end{equation}
Notice finally that for invertible $T^\pm$ eqs. (\ref{rtqyb11}--\ref{rtqyb44}) 
are equivalent to 
\be
S_{12}(\chi_1,\chi_2)\, R^+_2(\chi_2) = 
\nonumber \\
R^+_2(\chi_2)\, S_{12}(\chi_1,-\chi_2)\, , 
\label{rs1}
\end{equation}
\be
{\widetilde S}_{\,  12}(\chi_1,\chi_2)\, R^-_2(\chi_2) = 
\nonumber \\
R^-_2(\chi_2)\, {\widetilde S}_{12}(\chi_1,-\chi_2) \, . 
\label{rs2}
\end{equation}

Type I scattering data are remarkable because eqs. (\ref{rqyb11}--\ref{rtqyb44}) represent a 
meeting point between the existing approaches to factorized scattering 
with impurities and allow to compare them (see section 5).  

\bigskip 
\noindent{\bf Explicit type I solutions}:  An example \cite{Mintchev:2002zd} is  given by
the $GL(n)$--invariant scattering matrices
\be
S_{12}(\chi_1,\chi_2)={\widetilde S}_{21}(\chi_2,\chi_1)=
\frac{[s(\chi_1)-s(\chi_2)]\,\II\otimes\II-ig\,P 
_{12}}{s(\chi_1)-s(\chi_2)+ig} \, ,
\qquad g\in \RR\, ,
\label{IS11}
\end{equation} 
where $P_{12}$ is the standard flip operator and $s$ is an {\it even} real--valued function.
In this case, the complete classification of reflection and transmission matrices
is given by 
\begin{equation}
\quad R^\pm(\chi )  = \cos\left [\theta(\chi)\right ]\,
\exp [ip(\chi)\pm i\,n(\chi)]\,  \II \, , \qquad \quad 
\label{IR}
\end{equation}
\begin{equation}
T^{\pm}(\chi )=\sin\left [\theta(\chi)\right ]\, \exp[\pm iq(\chi)\pm
i\,n(\chi)]\, U^{\pm1} \, ,
\label{IT}
\end{equation}
where $U$ is a unitary matrix ($U^{-1}=U^\dag$),
$q$ is {\it even} and $\theta$, $p$ and $n$ are {\it odd}
real--valued functions. The parity of $s$ implies that
$S$ is not Lorentz (Galilean) invariant.
According to \cite{Castro-Alvaredo:2002fc}, 
this is a general feature of the type I solutions, following from (\ref{rs1}, \ref{rs2}).

\bigskip 
(ii) {\bf Type II} scattering data are characterized by setting 
\be
S^{++}(\chi_1,\chi_2) =  S^{+-}(\chi_1,\chi_2) =
S^{-+}(\chi_1,\chi_2) =
S^{--}(\chi_1,\chi_2) = S(\chi_1, \chi_2)\,  , 
\label{IIS}
\end{equation}
where $S$ obeys the QYBE. As a consequence, eqs.(\ref{qyb1}-\ref{qyb4}) are 
also satisfied and one is left with eqs. (\ref{rqyb1}-\ref{rtqyb2}) and their 
mirror counterparts. A general class of solutions of all these equations 
is given by the following matrices:
\be
R^\pm(\chi )= \cos\left [\theta(\chi)\right ]\, \exp [ ip_{\pm}(\chi)]\, 
B(\chi ) \, ,
\end{equation}
\be
T^\pm(\chi)=\sin\left [\theta(\chi)\right ]\, \exp[\pm iq(\chi)]\, 
\II\, ,
\end{equation}
where $\theta$, $p_{\pm}$ and $q$ are odd real--valued functions and
$B$ is a solution of the reflection QYBE relative to $S$ and satisfies eqs.
(\ref{runit}) and (\ref{rha}). This general procedure for deriving RT algebra 
scattering data $\{\cS,\, \R,\, \T\}$
from purely reflecting data $\{S,\, B\}$ is very attractive because 
of the large amount of existing results concerning the doublet $\{S,\, B\}$.

\bigskip 
\noindent{\bf Explicit type II solutions}: Taking for instance
\be
S_{12}(\chi_1,\chi_2)=\frac{(\chi_1-\chi_2)\,\II\otimes\II-ig\,P_{12}} 
{\chi_1-\chi_2+ig} \, , \qquad g\in \RR\, ,
\label{IIS1}
\end{equation} 
the general solution of (\ref{rqyb1}--\ref{rtqyb2}) can be parametrized according to 
\begin{equation}
R^\pm(\chi ) = \cos\left [\theta(\chi)\right ]\,
\exp [ip(\chi)\pm i\,n(\chi)]\,  \frac{\II+ia \chi \, U\, \EE\, U^{\dagger}}{1+ia 
\chi}\, ,  \qquad a\in \RR\, ,
\label{IIR}
\end{equation}
\begin{equation}
T^{\pm}(\chi )=\sin\left [\theta(\chi)\right ]\,
\exp[\pm iq(\chi)\pm i\,n(\chi)]\, \II \, ,
\label{IIT}
\end{equation}
where $\EE$ is a diagonal matrix which squares to $\II$ and U, 
$\theta$, $p$, $q$ and $n$ have the properties fixed in point (i).

The striking feature of type II data, which is manifest in the example (\ref{IIS1}-\ref{IIT}), 
is the coexistence of non-trivial transmission
with non--constant Lorentz (Galilean) invariant bulk scattering matrix.
This is clearly possible because the data violate 
(\ref{rtqyb11}--\ref{rtqyb44}), in spite of the fact that
all (\ref{qyb1}--\ref{rtqyb2}) are respected.

\bigskip 
(iii) {\bf Type III} scattering data are obtained by taking
\be
S^{++}(\chi_1,\chi_2) =  S^{+-}(\chi_1,-\chi_2) =
S^{-+}(-\chi_1,\chi_2) =
S^{--}(-\chi_1,-\chi_2) = S(\chi_1, \chi_2)\, ,
\label{IIIS}
\end{equation}
where $S$ is a solution of the QYBE.
The signs of the arguments in (\ref{IIIS}) are crucial for satisfying 
(\ref{qyb2}--\ref{qyb4})
and their mirror images. For deriving $\R$ and $\T$ one can proceed following 
the idea in point (ii). Take a solution $B$ of the reflection 
QYBE relative to $S$ and obeying eqs.
(\ref{runit}) and (\ref{rha}) from the case of pure reflection. Then,
using the properties of the previously introduced $\theta$, $p$ and 
$q$ functions, one can easily verify that
\be
R^\pm(\chi )= \cos\left [\theta(\chi)\right ]\, \exp [ ip(\chi)]\, 
B(\pm \chi ) \, ,
\label{IIIR}
\end{equation}
\be
T^\pm(\chi)=\sin\left [\theta(\chi)\right ]\, \exp[\pm iq(\chi)]\, 
B(\pm \chi)\, ,
\label{IIIT}
\end{equation}
satisfy (\ref{rha}, \ref{rtunit1}--\ref{tha}), all 
(\ref{rqyb1}--\ref{rtqyb2}) and their
mirror counterparts. 

Type III scattering data are inspired by the NLS model with point--like impurity,
which has non--trivial bulk scattering but is nevertheless
exactly solvable \cite{Caudrelier:2004gd}--\cite{CR}.  For the time being it
is unique with these properties and provides therefore
a valuable test for  the whole RT algebra framework. 
Since the NLS model is well--known to 
capture many of the universal properties of integrable system, we strongly 
believe that type III solutions are relevant also in a more general context. 

\bigskip 
\noindent{\bf Explicit type III solutions}: As already mentioned, 
the NLS  model provides an instructive example. The relative scattering data 
are parametrized by (\ref{IIIS}--\ref{IIIT}) with $S$ defined by (\ref{IIS1}) and $B$ 
given by \cite{Mintchev:2001aq}
\be
B(\chi ) = \exp [ i\,n(\chi)]\,\frac{\II+ia \chi \, U\, \EE\, U^{\dagger}}{1+ia \chi} \, , 
\qquad a \in \RR\, , 
\label{B}
\end{equation}
where $U$ and $n$ are defined as above.
Because of (\ref{IIIS}), $\cS$ is not Lorentz (Galilean) invariant in spite of the fact 
that $S$ preserves this symmetry. We refer to \cite{Caudrelier:2004gd}--\cite{CR} for more details. 
\bigskip

\bigskip 

\sect{Remarks and conclusions} 

As already mentioned, the results of the present paper allow to establish a direct 
relationship between the RT algebra framework and the approach previously 
developed in \cite{Cherednik:1990ju}-\cite{Delfino:1994nr}. 
Extending his idea about purely reflecting boundaries 
(mirrors in his terminology), Cherednik introduced in \cite{Cherednik:1990ju, Cherednik:jt} the concept 
of purely transmitting defects (glasses). These ideas have been later independently generalized for 
impurities which both reflect and transmit by 
Delfino, Mussardo and Simonetti (DMS) in \cite{Delfino:1994nr}. 
The scattering data of the DMS approach consists of a bulk
scattering  matrix $S$ and right (left) reflection and transmission matrices $R^+$ ($R^-$) and $T^+$ ($T^-$). 
As expected, $S$ must satisfy the QYBE. Remarkably enough, the consistency relations 
imposed in \cite{Delfino:1994nr} on $R^\pm$ and $T^\pm$, 
precisely coincide with the special case (\ref{rqyb11}--\ref{rtqyb44}) of type I 
scattering data, provided that 
\be 
{\widetilde S}_{12}(\chi_1,\chi_2) = S_{21}(\chi_2,\chi_1)\, , 
\label{Stilde}
\end{equation} 
which, as easily verified, solves the QYBE as well. We conclude therefore that 
the systems discussed in \cite{Delfino:1994nr} are fully described by type I data 
in the RT algebra framework. RT algebras are however 
significantly more general in the sense that they allow for a larger set of 
scattering data, still leading to a unitary scattering operator \cite{Mintchev:2003ue}. 
In fact, both type II and type III solutions are not covered by the DMS 
approach. That is why the physical example of the NLS model with impurity 
\cite{Caudrelier:2004gd}--\cite{CR} can not be treated along the lines 
of \cite{Delfino:1994nr}: the corresponding scattering data 
satisfy the RT algebra reflection--transmission QYBE, but do not
respect the DMS  consistency conditions. 

Summarizing, the concept of RT algebra has a richer structure and thus opens 
new possibilities. Among others, we would like to mention the 
type II solutions with both Lorentz (Galilean) invariant non-constant bulk 
scattering matrix and non-vanishing transmission matrix, which 
are forbidden \cite{Castro-Alvaredo:2002fc} in the framework 
of \cite{Delfino:1994nr}. It will be interesting in this respect to 
construct exactly solvable models, possessing such scattering data. 
Some recent developments in model--building with impurities can be found in 
\cite{Bowcock:2004my}--\cite{HLP}. 

In conclusion, the RT algebra approach to integrable systems with impurities is 
based on the reflection--transmission QYBE's (\ref{rqyb}, \ref{tqyb}, \ref{rtqyb}). 
Keeping in mind that (\ref{tqyb}, \ref{rtqyb}) are deeply related to (\ref{rqyb}) 
by physical unitarity, this beautiful and compact set of equations is in our opinion extremely natural 
and deserves further investigation.

\bigskip
\bigskip 
{\bf Acknowledgments:} M.~M. would like to thank l'Universit\'e de Savoie for 
the financial support and LAPTH in Annecy for the kind hospitality. V.~C. kindly 
acknowledges the financial support from INFN and
the warm hospitality of the theory group in Pisa. 
Work supported in part by the TMR Network EUCLID: ``Integrable models and
applications:  from strings to condensed matter", contract number HPRN-CT-2002-00325.


\begin{thebibliography}{99} 



\bibitem{Yang:1967bm}
C.~N.~Yang,
Phys.\ Rev.\ Lett.\  {\bf 19} (1967) 1312.

\bibitem{Sklyanin:ye}
E.~K.~Sklyanin,
Sov.\ Phys.\ Dokl.\  {\bf 24} (1979) 107
[Dokl.\ Akad.\ Nauk Ser.\ Fiz.\  {\bf 244} (1978) 1337].

\bibitem{Karowski:1978vz}
M.~Karowski and P.~Weisz,
Nucl.\ Phys.\ B {\bf 139} (1978) 455.

\bibitem{Zamolodchikov:xm}
A.~B.~Zamolodchikov and A.~B.~Zamolodchikov,
Ann. Phys. (N.Y.)\  {\bf 120} (1979) 253.

\bibitem{Faddeev:gh}
L.~D.~Faddeev, E.~K.~Sklyanin and L.~A.~Takhtajan,
Theor.\ Math.\ Phys.\  {\bf 40} (1980) 688
[Teor.\ Mat.\ Fiz.\  {\bf 40} (1979) 194].

\bibitem{Grosse} H.~Grosse, 
Phys.\ Lett.\ B {\bf 86} (1979) 267. 

\bibitem{Honerkamp:1979mx}
J.~Honerkamp, P.~Weber and A.~Wiesler,
Nucl.\ Phys.\ B {\bf 152} (1979) 266.

\bibitem{Creamer:1979an}
D.~B.~Creamer, H.~B.~Thacker and D.~Wilkinson,
Phys.\ Rev.\ D {\bf 21} (1980) 1523.

\bibitem{Faddeev:zy}
L.~D.~Faddeev,
Sov.\ Sci.\ Rev.\ C {\bf 1} (1980) 107.

\bibitem{Davies:gc}
B.~Davies,
J.\ Phys.\ A {\bf 14} (1981) 2631.

\bibitem{Liguori:de}
A.~Liguori, M.~Mintchev and M.~Rossi,
J.\ Math.\ Phys.\  {\bf 38} (1997) 2888.

\bibitem{Cherednik:vs}
I.~V.~Cherednik,
Theor.\ Math.\ Phys.\  {\bf 61} (1984) 977
[Teor.\ Mat.\ Fiz.\  {\bf 61} (1984) 35].

\bibitem{Sklyanin:yz}
E.~K.~Sklyanin,
J.\ Phys.\ A {\bf 21} (1988) 2375.

\bibitem{Kulish:pb}
P.~P.~Kulish and R.~Sasaki,
Prog.\ Theor.\ Phys.\  {\bf 89} (1993) 741
[arXiv:hep-th/9212007].

\bibitem{Ghoshal:tm}
S.~Ghoshal and A.~B.~Zamolodchikov,
Int.\ J.\ Mod.\ Phys.\ A {\bf 9} (1994) 3841
[Erratum, ibid.\ A {\bf 9} (1994) 4353]
[arXiv:hep-th/9306002].

\bibitem{Corrigan:1994np}
E.~Corrigan, P.~E.~Dorey and R.~H.~Rietdijk,
Prog.\ Theor.\ Phys.\ Suppl.\  {\bf 118} (1995) 143
[arXiv:hep-th/9407148].

\bibitem{Fring:1993mp}
A.~Fring and R.~Koberle,
Nucl.\ Phys.\ B {\bf 421} (1994) 159
[arXiv:hep-th/9304141].

\bibitem{Liguori:1998xr}
A.~Liguori, M.~Mintchev and L.~Zhao,
Commun.\ Math.\ Phys.\  {\bf 194}, 569 (1998)
[arXiv:hep-th/9607085].

\bibitem{A} S. Albeverio, F. Gesztesy, R. Hoegh-Krohn ans H. Holden,
Solvable models
in quantum mechanics (Springer-Verlag, New York, 1988).

\bibitem{Mintchev:2002zd}
M.~Mintchev, E.~Ragoucy and P.~Sorba,
Phys.\ Lett.\ B {\bf 547} (2002) 313
[arXiv:hep-th/0209052].

\bibitem{Mintchev:2003ue}
M.~Mintchev, E.~Ragoucy and P.~Sorba,
J.\ Phys.\ A {\bf 36} (2003) 10407
[arXiv:hep-th/0303187].

\bibitem{G} 
F. R. Gantmacher, The Theory of Matrices, vol. 1 (Chelsea, New York, 1959). 

\bibitem{H} N. J. Higham, Linear Algebra Appl. 88/89 (1987) 405. 

\bibitem{Mintchev:2003kh}
M.~Mintchev and E.~Ragoucy, 
J.\ Phys.\ A {\bf 37} (2004) 425
[arXiv:math.qa/0306084].

\bibitem{Ragoucy:2001cf}
E.~Ragoucy, 
Lett.\ Math.\ Phys.\  {\bf 58} (2001) 249
[arXiv:math.qa/0108221].

\bibitem{Caudrelier:2004gd}
V.~Caudrelier, M.~Mintchev and E.~Ragoucy,
J.\ Phys.\ A {\bf 37} (2004) L367
[arXiv:hep-th/0404144].

\bibitem{Caudrelier:2004xy}
V.~Caudrelier, M.~Mintchev and E.~Ragoucy, 
``Solving the quantum non-linear Schrodinger equation with
delta-type impurity,'' arXiv:math-ph/0404047, J.\ Math.\ Phys.\ in press. 

\bibitem{CR} 
V.~Caudrelier and E.~Ragoucy, 
``Spontaneous symmetry breaking in the non--linear Schr\"odinger hierarchy with defect", 
arXiv:math-ph/0411022.  

\bibitem{Mintchev:2004jy}
M.~Mintchev and P.~Sorba,
JSTAT {\bf 0407} (2004) P001
[arXiv:hep-th/0405264].

\bibitem{Freidel:1991jx}
L.~Freidel and J.~M.~Maillet, 
Phys.\ Lett.\ B {\bf 262} (1991) 278.

\bibitem{Cherednik:1990ju}
I.~Cherednik, ``Notes on affine Hecke algebras. 1. Degenerated affine Hecke 
algebras and Yangians in mathematical physics,'' BONN-HE-90-04. 

\bibitem{Cherednik:jt}
I.~Cherednik,
Int.\ J.\ Mod.\ Phys.\ A {\bf 7} (1992) 109.

\bibitem{Delfino:1994nr}
G.~Delfino, G.~Mussardo and P.~Simonetti,
Nucl.\ Phys.\ B {\bf 432} (1994) 518
[arXiv:hep-th/9409076].

\bibitem{Bajnok:2004jd}
Z.~Bajnok and A.~George, 
``From defects to boundaries,''
arXiv:hep-th/0404199.

\bibitem{Castro-Alvaredo:2002fc}
O.~A.~Castro-Alvaredo, A.~Fring and F.~Gohmann,
On the absence of simultaneous reflection and transmission in
integrable impurity systems,
arXiv:hep-th/0201142.

\bibitem{Mintchev:2001aq}
M.~Mintchev, E.~Ragoucy and P.~Sorba, 
J.\ Phys.\ A {\bf 34} (2001) 8345
[arXiv:hep-th/0104079].

\bibitem{Bowcock:2004my}
P.~Bowcock, E.~Corrigan and C.~Zambon, 
JHEP {\bf 0401} (2004) 056
[arXiv:hep-th/0401020].

\bibitem{Corrigan:2004se}
E.~Corrigan and C.~Zambon, 
J.\ Phys.\ A {\bf 37} (2004) L471
[arXiv:hep-th/0407199].

\bibitem{Hallnas:2004dn}
M.~Halln\"as and E.~Langmann,
``Exact solutions of two complementary 1D quantum many-body systems on the half-line,''
arXiv:math-ph/0404023.

\bibitem{HLP} M.~Halln\"as, E.~Langmann and C.~Paufler. 
``Generalized local interactions in 1D: solutions of quantum many-body systems 
describing distinguishable particles", 
arXiv:math-ph/0408043.

\end{thebibliography}
\end{document}